\documentclass[review]{elsarticle}

\usepackage{graphicx,hyperref}
\usepackage{amsmath,amsthm}
\newtheorem{theorem}{Theorem}
\theoremstyle{definition}
\newtheorem{definition}{Definition}

\journal{Information Processing Letters}

\usepackage{xcolor}

\begin{document}

\begin{frontmatter}

\title{NP-completeness of the Active Time Scheduling Problem}

\author{Sagnik Saha}
\address{Google, Mountain View}
\ead{sagnik1saha@gmail.com}
\author{Manish Purohit}
\address{Google Research, Mountain View}
\ead{mpurohit@google.com}

\begin{abstract}
In this paper, we study the active time scheduling problem. We are given \(n\) jobs with integral processing times each of which has an integral release time and deadline. The goal is to schedule all the jobs on a machine that can work on \(b\) jobs simultaneously, and the objective is to minimize the number of time slots for which the machine is active.
The active time scheduling model was introduced by Chang et al.\ (Algorithmica, 2015) in the context of energy-efficient scheduling. Surprisingly, despite the development of a number of constant factor approximation algorithms for the problem, the complexity of this fundamental problem had remained open. In this paper, we resolve this open problem and show that the active time scheduling problem is indeed NP-complete.

\end{abstract}


\end{frontmatter}

\section{Introduction}

Energy efficient scheduling is a fundamental problem that has received attention in the computer science and operations research communities for decades. Fuelled by the widespread adoption of cloud computing and the availability of energy-efficient hardware that enables slowing down processor speeds and turning off unused processors,
minimizing energy consumption by consolidating load on a few active machines at a time has emerged as a leading approach for efficient scheduling~\cite{beloglazov2010energy,zhang2010cloud,goiri2010energy}. In the \emph{active time scheduling} model introduced by Chang et al.\ \cite{chang2012model}, the machine can be turned off when it is not in use and can be used to schedule up to $b$ jobs at a time if it is active. Each job $i$ has a processing demand $p(i)$ and an interval $[r(i), d(i)]$ between which it can be scheduled. The goal is to find a schedule to execute all the jobs while minimizing the energy consumption, i.e., the total number of time slots when the machine is active.

The \emph{active time scheduling} model has been well-studied~\cite{chang2012model,chang2017lp,kumar2018brief,calinescu2021new} and a number of approximation algorithms are known. In particular, Chang et al.\ \cite{chang2017lp} and Calinescu et al.\ \cite{calinescu2021new} give a 2-approximation algorithm based on rounding the natural linear programming relaxation. While Chang et al.\ \cite{chang2017lp} show that any minimal feasible solution yields a 3-approximation, Kumar et al.\ \cite{kumar2018brief} demonstrate that a simple greedy algorithm that shuts off unnecessary time slots from left to right also yields a 2-approximation. Despite the wealth of approximation algorithms, surprisingly the complexity of this fundamental scheduling problem remained open\footnote{Both Kumar et al.\ \cite{kumar2018brief} and Calinescu et al.\ \cite{calinescu2021new} explicitly state resolving the time complexity as open questions.}. In this paper, we resolve this question and show that the \emph{active time scheduling} problem is indeed NP-hard.

\begin{theorem}\label{Theorem:np-complete}
The \emph{active time scheduling} problem is NP-complete.
\end{theorem}

The proof uses a reduction from a natural variant of the \emph{boolean satisfiability} problem and requires a delicate construction to maintain a mapping between active machine times and boolean values for the variables. We formally describe the \emph{active time scheduling} problem in Section \ref{sec:problem}, present the reduction in Section \ref{sec:reduction}, and finally present the proof in Section \ref{sec:proof}.

\subsection{Other Related Work}
A variant of the active time scheduling problem where each job has a \emph{set} of time intervals when it can be scheduled is known to be NP-complete when the batch size $b \geq 3$~\cite{chang2012model}. On the other hand, if all jobs have a unit processing demand, then the problem can be solved in polynomial time \cite{chang2012model}. Similarly, when the batch size $b$ is unbounded, the problem reduces to the hitting set problem on intervals and hence admits a polynomial time solution.

\section{Model and Preliminaries}
\label{sec:problem}

We first formally define the \emph{Active Time Scheduling} problem.

\begin{definition}[Active Time Scheduling Problem]
We are given \(n\) jobs with integer processing times \(p_1, p_2,..., p_n\) respectively. We assume that time is slotted and each job $i$ has an integer release time \(r_i\) and a deadline \(d_i\); the \(i^{th}\) job must be scheduled for \(p_i\) time slots within the interval \([r_i, d_i]\). We assume that preemption is allowed, and a job can be preempted at integral time steps. Finally, for a fixed batch size $b$, the machine can run up to $b$ distinct jobs at each time slot if it is active.

Our objective is to schedule all the jobs for the appropriate processing times within their corresponding intervals in order to minimize the number of time slots when the machine is active (i.e., when at least one job is scheduled). In the decision version of the problem, we are also given a target cost \(t\), and we are asked whether it’s possible to schedule all the jobs under the given constraints with at most \(t\) active time slots being used.
\end{definition}

We will prove that the active time scheduling problem is NP-complete. It is easy to see that it’s in NP, since a complete schedule takes linear time to verify and hence, can serve as a valid certificate if the answer is yes. To prove that it’s NP-hard, we’ll reduce from the \emph{Balanced SAT} problem defined below. %

\begin{definition}[Balanced SAT]
In Balanced SAT, we are given a boolean formula $F(x_1, x_2, ..., x_n)$ where the number of variables, $n$, is even. $F$ is presented in CNF (AND of ORs) where each clause is an OR of several literals. The problem asks whether there is a satisfiable assignment of the boolean variables that sets exactly $n/2$ variables to $true$.
\end{definition}

This problem is NP-hard because of an easy reduction from SAT. Given a SAT formula \(F\) with \(n\) variables \(x_1, x_2, ..., x_n\), we construct a Balanced SAT instance as follows. Introduce \(n\) new variables \(y_1, y_2, ..., y_n\) and for each \(i \in [1, n]\),  introduce 2 new clauses \((x_i \vee y_i)\) and \((\overline{x_i} \vee \overline{y_i})\). The new clauses together enforce the constraint 
\begin{equation}
\label{BSAT-eqn}
y_i=\overline{x_i} \qquad \forall 1\le i \le n
\end{equation}
It's easy to check that this new formula is satisfiable if and only if the original formula is satisfiable. Furthermore, the new formula has an even number of variables, and Equation \eqref{BSAT-eqn} implies that any of its solutions has exactly half of the variables set to true.

\section{Reduction}
\label{sec:reduction}

Let $F$ be an arbitrary instance of the Balanced SAT problem, where the number of clauses is \(m\), and the number of variables is \(n\). Assume that the $k^{th}$ clause contains \(n_k\) literals. In our construction, we’ll set the target cost $t$, and the batch size $b$ as follows:
\begin{equation}t=m+2+\sum_{k=1}^mn_k\end{equation}
and
\begin{equation}b = 2n + 2\end{equation}

\begin{figure}
    \centering
    \includegraphics[width=\textwidth,page=1]{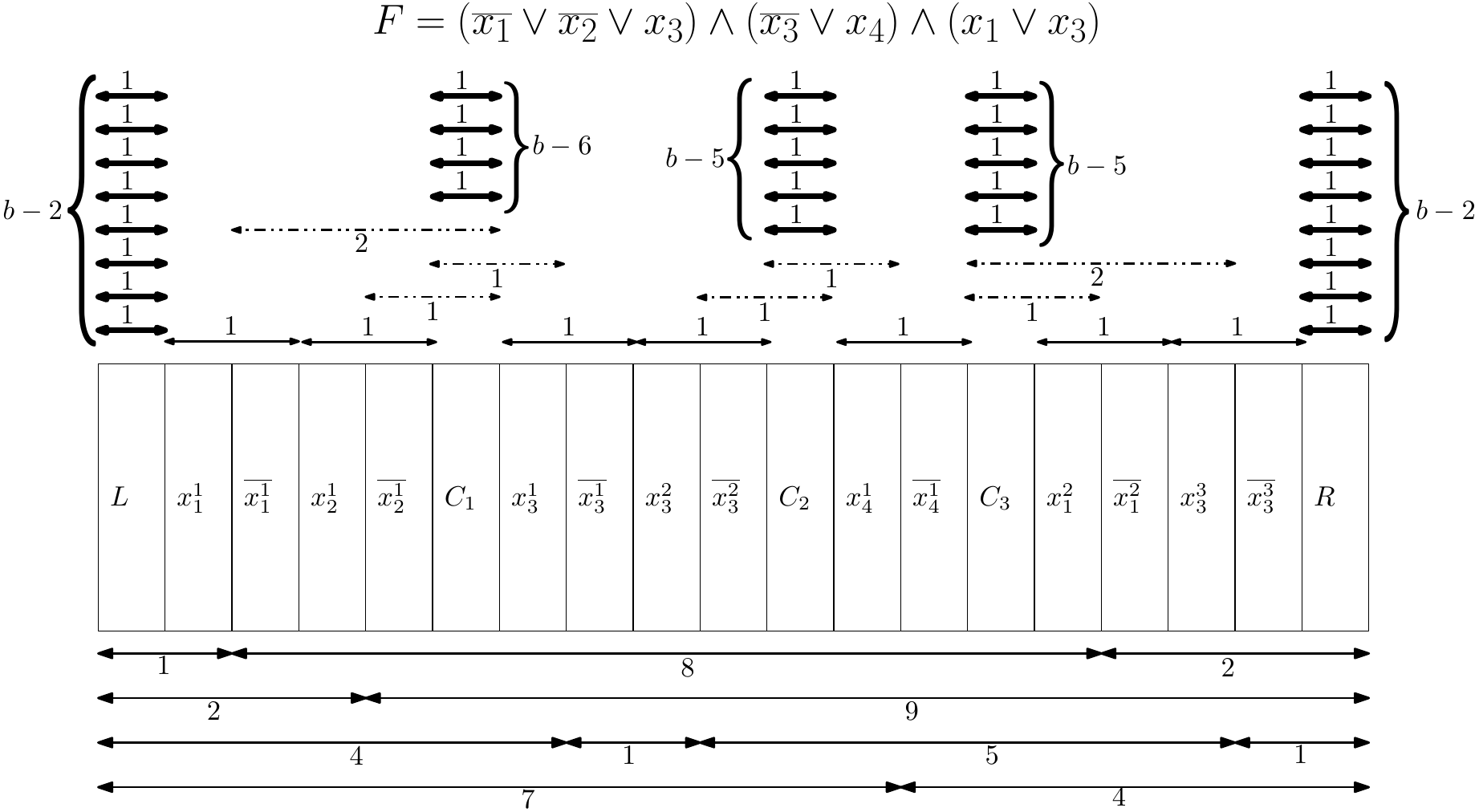}
    \caption{A typical translation of a SAT instance. Each rectangle represents one time slot. The different arrows all represent jobs, with the two endpoints corresponding to its release time and the deadline, and the associated number representing its processing time. Since the formula has 4 variables, the value of \(b =2\cdot 4+2=10\). The arrows are shaded differently to show different types of jobs in our reduction. All the bold arrows correspond to jobs with an active interval of size 1 and a processing time of 1; their purpose is to reduce the effective batch size in those time slots. The solid arrows at the top correspond to variable gadgets; they are meant to induce a binary choice representing the two possible assignments of the variable. The dashed arrows are a part of the clause gadgets; they ensure that at least one literal in each clause satisfies it. The solid arrows at the bottom correspond to copy gadgets; these jobs ensure that every copy of the same variable is set the same way.}
    \label{whole_picture}
\end{figure}

For ease of description, we name the time units as shown in Fig. \ref{whole_picture} (instead of \(0, 1, 2, ...\)). The leftmost time slot is called \(L\) . This is followed by several clause gadgets, and then the last time slot is called \(R\). Each clause gadget consists of several variable gadgets (corresponding to all the literals in that clause) and one extra time slot, as shown. We’ll describe each of these gadgets in detail below. 

\subsection{Variable gadget}
One variable can be used in several different clauses. We denote the variable index by the subscript, and the superscript indicates which instance (or copy) of the variable we are talking about.

For the \(j\)th instance of the variable \(x_i\), we create a job with processing time 1 which has to be completed within an interval of 2 time units as shown above. The corresponding release time is \(x_i^j\) and the corresponding deadline is \(x_i^j+1=\overline{x_i^j}\).

Note that this forces a choice upon the solver. If the solver schedules this job in the second available time unit (marked \(\overline{x_i^j}\)), we interpret that as setting the variable \(x_i\) to be \(false\). If the job is scheduled on the first available spot, \(x_i\) is taken to be \(true\). Our copy gadget ensures we set the variable consistently in all the instances; we’ll describe it later.

\begin{figure}
    \centering
    \includegraphics[scale=1,page=2]{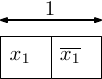}
    \caption{A variable gadget.}
    \label{variable}
\end{figure}

\subsection{Clause gadget}
Let’s consider a typical clause: \[C_k = (\overline{x_{a_1}} \vee \overline{x_{a_2}} \vee ... \vee \overline{x_{a_p}} \vee x_{a_{p+1}} \vee ... \vee x_{a_{n_k}})\]
We’ll first separate all the positive literals from the negative literals, which we’ve already done in the above canonical form. Going from left to right, we first place one variable gadget corresponding to each of the negative literals (the instance number depends on how many earlier clauses used that variable). Next, we place the time slot called \(C_k\). Next, we place one variable gadget for each of the positive literals.

Each variable gadget already contains 1 job, as described before. We create \(b-n-n_k+1\) jobs with processing time 1 which have to be scheduled at time \(C_k\) (they have both release time and deadline set to \(C_k\)). 

Next, for each variable gadget included in this clause gadget, we create a job stretching from \(C_k\) to the center of that gadget. The processing time of that job is set to be 1 more than the number of full variable gadgets this job spans over.

Formally speaking, we create \(n_k\) jobs, one for each variable in the clause. In the canonical clause above, we assumed that there are \(p\) negative literals. As long as \(q\le p\), we create a job with release time \(\overline{x_{a_q}}\), deadline \(C_k\), and processing time \(p-q+1\). If \(q>p\), we create a job with release time \(C_k\), deadline \(x_{a_q}\), and processing time \(q-p\).
Note that we skipped the superscripts above for the sake of legibility; we are always referring to the instance of each variable that’s included in the clause gadget under discussion.

\begin{figure}
    \centering
    \includegraphics[page=3]{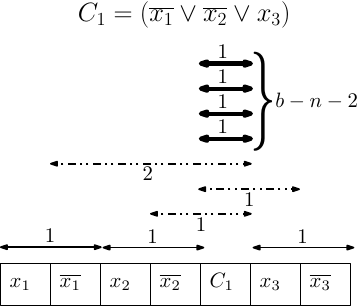}
    \caption{A clause gadget}
    \label{clause}
\end{figure}

\subsection{Copy gadget}
We have to ensure that each instance of a variable is set the same way (i.e., either they are all set to \(true\) or all set to \(false\)). Consider a typical variable \(x_i\). Let’s assume it has \(r\) instances in total. Our copy gadget will consist of \(r+1\) jobs. For each \(1\le j<r\), we create a job with release time \(\overline{x_i^j}\) and deadline \(x_i^{j+1}\). We also create one job with release time \(L\) and deadline \(x_i^1\), and one job with release time \(\overline{x_i^r}\) and deadline \(R\).

These jobs can all have different processing times. We determine the processing time of any of these jobs in the same fashion. We count the total number of full variable gadgets and the number of clause time slots the job spans over, and set its processing time to \(1 + \textrm{that value}\).

\subsection{Handling L and R}
We schedule \(b-n/2\) jobs of processing time 1 that have to be scheduled at \(L\) (i.e. their release times and deadlines are all set to \(L\)). We do the same thing for \(R\).

\section{Proof}
\label{sec:proof}

 We have to show that this instance of the scheduling problem is feasible if and only if the original Balanced SAT problem is solvable. We’ll prove both directions below.

\subsection{Balanced SAT solution to Scheduling solution}
Let’s start with a satisfying boolean assignment.

\paragraph{Step 1}\label{Step 1} For the sake of simplicity, start by scheduling all the jobs with unit interval length (release time = deadline) in the only time slot eligible for them.

\paragraph{Step 2}\label{Step 2} For each variable, schedule all the jobs corresponding to all of its instances on the left slots if the corresponding variable is true, and schedule them on the right slots otherwise. So, if there are \(r\) instances, we’ll schedule a job at \(x_i^j\) for all \(1\le j\le r\) if the variable \(x_i\) was set to \(true\), and otherwise those slots will remain empty and we’ll schedule a job at \(\overline{x_i^j}\) for all \(1\le j\le r\). 

Note that at this point, we already have \(t\) active time slots. \(L\) and \(R\) as well as all the \(C_k\) slots were activated in the first step since we have unit interval jobs on those slots, creating \(m+2\) active time slots. In the second step, we activated one time slot per instance of each variable, creating another \(\sum_{k=1}^mn_k\) active time slots. In the rest of the translation, we will not create any new active time slot. So slots which are empty at this stage will always remain empty.

\paragraph{Step 3}\label{Step 3} For each clause \(C_k\), we must have at least one literal satisfying it. Assume it’s a negative literal \(\overline{x_i}\). Then, schedule the job corresponding to that variable in all the active time slots spanned by this job except \(C_k\). This means that if \(C_k\) has \(p\) negated literals, we’ll schedule the job in \(\overline{x_i}\) and in all the \(p-i\) variable gadgets following it, thus satisfying the processing time of \(p-i+1\). For all the other \(n_k-1\) jobs corresponding to the other variables, schedule them in \(C_k\) and in all of the full variable gadgets they span over.

We do essentially the same thing if \(C_k\) is satisfied by a positive literal \(x_i\). The job corresponding to that variable gets scheduled in the slot \(x_i\) as well as the \(i-p-1\) variable gadgets preceding it, thus satisfying the processing time if \(i-p\). All the other jobs in the gadget are scheduled exactly as before.

\paragraph{Step 4}\label{Step 4} Now we schedule all the jobs for the copy gadgets. The jobs that don’t span either \(L\) or \(R\), start at \(\overline{x_i^j}\) and end at \(x_i^{j+1}\), exactly one of which is active. All the clause time slots they span are active, and exactly one slot is active per full variable gadget they span over. Therefore, their processing time is exactly equal to the number of active slots in their interval, and we simply schedule them in all the active time slots in their interval.

If \(x_i\) is false, we’ll have a similar situation for the job starting at \(L\) and ending at \(x_i^1\), and we simply schedule that job in all the active time slots in its interval (including \(L\)). For the job starting at \(\overline{x_i^r}\) (where \(r\) is the number of instances of \(x_i\) in our formula) and ending at \(R\), we don’t use \(R\). Since \(\overline{x_i^r}\) is active, we’ll again be able to schedule the job completely using the \(\overline{x_i^r}\) slot and all the active slots between that and \(R\). We do just the opposite thing for true variables, i.e., we use the \(R\) slot and avoid the \(L\) slot.

\paragraph{Verification}\label{Verification} Our construction so far clearly respects the release time, deadline, and the processing time for each job. Since we had \(t\) active time slots at the end of \nameref{Step 2} and didn’t activate any time slot since then, we incur a cost of exactly \(t\). We now just need to check that the batch size \(b\) is respected everywhere.

\begin{itemize}
\item The \(L\) slot is used by \(b-n/2\) unit interval jobs in \nameref{Step 1}, and once for each false variable in \nameref{Step 4}. Since we know that exactly \(n/2\) variables are false, that means this slot is used by exactly \(b\) jobs. By a symmetric argument, we can also show that the \(R\) slot also respects the batch size.
\item Consider any instance of a variable gadget. It’s used once to set the variable in \nameref{Step 2}. It can be used by several of the large jobs corresponding to the clause gadget it belongs to in \nameref{Step 3}. Since a clause contains at most \(n\) literals, we use this gadget at most \(n\) times in \nameref{Step 3}. And lastly, since the copy gadget jobs corresponding to any one variable are non-overlapping and there are \(n\) variables, we use this gadget a further \(n\) times in \nameref{Step 4}. Therefore, the total number of jobs scheduled in the one active time slot in this gadget is bounded above by \(1+n+n<b\).
\item Finally, consider a clause time slot \( C_k\). It’s used by \(b-n-n_k+1\) unit interval jobs in \nameref{Step 1}. In \nameref{Step 4}, the copy gadget for each variable uses this exactly once, thus adding another \(n\) jobs to this slot. In \nameref{Step 3}, we used this slot for every literal in the clause except one, thus adding \(n_k-1\) jobs. Therefore, we have exactly \(b\) jobs scheduled in this time slot.
\end{itemize}

Hence we have a solution for the instance of our scheduling problem. 

\subsection{Scheduling solution to Balanced SAT solution}

Now let’s start with a valid schedule for all the jobs in our scheduling problem with cost at most \(t\), and derive a solution to the corresponding Balanced SAT problem instance.

\paragraph{Counting active slots} All the unit interval jobs can be scheduled in only one way, so we can safely assume that the \(L\) and \(R\) slots as well as all the \(C_k\) slots are active. Each of the variable instance gadgets have one job with unit processing time, and the intervals are all pairwise non-overlapping. So, at least one time slot must be active per variable gadget. Since the \(k^{th}\) clause gadget contains \(n_k\) variable gadgets, this accounts for another \(\sum_{k=1}^mn_k\) active slots.

Observe that we’ve already accounted for \(2+m+\sum_{k=1}^mn_k=t\) active time slots. Since the scheduling is a valid solution, we know that all the remaining time slots must be inactive. In particular, \emph{no variable gadget can have both of its slots active}.

\paragraph{Verifying the Copy gadget} For now, let’s pretend each instance of a variable is a separate variable. An instance is set to \(true\) if the left slot in the corresponding gadget is active, and it’s set to \(false\) otherwise. 

Consider a variable \(x_i\). Let’s assume it has \(r\) instances. For any $j \in [1, r-1]$, observe that this variable’s copy gadget includes a job with release time \(\overline{x_i^j}\) and deadline \(x_i^{j+1}\). If we have set \(x_i^j\) to \(true\) and \(x_i^{j+1}\) to \(false\), this job can’t be scheduled either on its release time or on its deadline. The number of active time slots available for this job then equals the number of clause time slots and variable gadgets it spans over. But by construction, its processing time is one more than this value, and we have a contradiction. Therefore, 
\begin{equation}\label{chain-impl}
    x_i^j \Rightarrow x_i^{j+1} \quad \forall 1\le j<r \quad \forall i
\end{equation}

For variable \(x_i\), consider the job with release time \(L\) and deadline \(x_i^1\). If \(x_i^1\) is \(false\), the number of active slots available for this job equals the number of clause time slots and variable gadgets it spans over, plus one (for the \(L\) slot). This matches the processing time of that job, and hence we are forced to schedule this job in each of those slots. In particular this shows that \emph{we need to use the \(L\) time slot for a copy gadget job corresponding to \(x_i\) if the first instance of that variable \(x_i^1\) is set to \(false\)}. Similarly, we can show that \emph{we need to use the \(R\) time slot for a copy gadget job corresponding to \(x_i\) if the last instance of that variable \(x_i^r\) is set to \(true\)}.

Now, for each variable, consider the first and last instance. Applying \autoref{chain-impl} in a chain we can show that if the first instance is true, the last instance can’t be false. So the possible truth value assignments of \(x_i^1\) and \(x_i^r\) are \(TT\), \(FF\) or \(FT\). Observe that in all cases, we have to use at least one of the \(L\) and \(R\) time slots for copy gadget jobs corresponding to each variable. However, these 2 slots have total capacity \(2b\), and unit interval jobs already take up \(2b-n\) of that capacity. Therefore, we only have space to use these two time slots a total of \(n\) times for all the copy gadget jobs corresponding to all the variables. This means that we can’t have \(x_i^1=F\) and \(x_i^r=T\) for any \(i\) (because then we’d need to use both the \(L\) and \(R\) time slots for copy gadget jobs corresponding to this one variable, and together with all the other variables which all require at least one use of either of the those two time slots, we’d need \(n+1\) space which we don’t have).

So we have established that the first and last instances of each variable are both set to true or both set to false. \autoref{chain-impl} now shows that all the instances of a variable are set to the same value.

We can therefore now talk about the value of a variable, instead of the value of a particular instance of a variable. We are going to show that setting the variables in this manner satisfies the original Balanced SAT problem.

\paragraph{Counting true variables} In  the last paragraph, we showed that if a variable is set to \(false\), we must use the \(L\) time slot for a copy gadget job corresponding to that variable. Since we can only use the \(L\) time slot \(n/2\) times for scheduling copy gadget jobs, this shows that we can have at most \(n/2\) \(false\) variables. Similarly we can have at most \(n/2\) \(true\) variables. Since the total number of variables is \(n\), both of these bounds must be tight, and \emph{exactly half of the variables are \(true\) and the other half are \(false\)}.

\paragraph{Proving the Clause gadget} We claim that each \(C_k\) time slot is used exactly once by the copy gadget jobs corresponding to any particular variable \(x_i\). The copy gadget jobs for any one variable are always non-overlapping, and together cover the whole timeline. Therefore, our claim is equivalent to stating that each copy gadget job uses all the clause time slots in its interval.

To prove this, we consider two cases. Here, we assume that \(x_i=T\). The proof works very similarly if the variable is \(false\) instead.
\begin{itemize}
\item First, consider a copy gadget job starting at \(\overline{x_i^j}\) and ending at \(x_i^{j+1}\). We know that exactly one slot among the release time and deadline is active (since the instances of a variable are consistent with each other). Therefore by construction, the number of active time slots available for this job equals its processing time. So we have no choice but to schedule this job at each active time slot in its interval, including all the clause time slots there.
\item The copy gadget job starting at \(\overline{x_i^r}\) (\(r\) is the number of instances of \(x_i\)) and ending at \(R\) also has a similar constraint. The release time is inactive since \(x_i=T\). By construction, the number of active time slots available for this job equals its processing time and we have to use all the clause time slots in its interval for this job.
\item The copy gadget job starting at \(L\) and ending at \(x_i^1\) can’t be scheduled at \(L\). This is because we have to use the \(L\) time slot for every \(false\) variable, of which there are \(n/2\). Since that slot also has \(b-n/2\) unit interval jobs, we don’t have any space there to schedule a copy gadget job corresponding to a positive variable like \(x_i\). Hence, we will be forced to use all the other active time slots available in the interval for this job, including \(x_i^1\) and all the clause time slots in this interval.
\end{itemize}

It now follows that \emph{for any \(k\), the copy gadget jobs together use the time slot \(C_k\) exactly \(n\) times}.

\paragraph{Deriving the SAT solution} We’ll use a proof by contradiction to show that each clause in our SAT formula is satisfied. Assume that the \(k^{th}\) clause is not satisfied. Then, all the positive literals in the clause are set to \(false\) and the negative literals are all set to \(true\). We assume, as before, that \[C_k = (\overline{x_{a_1}} \vee \overline{x_{a_2}} \vee ... \vee \overline{x_{a_p}} \vee x_{a_{p+1}} \vee ... \vee x_{a_{n_k}})\] 
\begin{itemize}
\item A typical job corresponding to \(x_{a_q}\), as long as \(q\le p\), has release time \(\overline{x_{a_q}}\), deadline \(C_k\), and processing time \(p-q+1\). Since we assumed that \(\overline{x_{a_q}}\) is inactive, we have exactly \(p-q\) active time slots from variable gadgets and 1 active time slot at \(C_k\) in the interval. As the number of active slots in the job’s interval equals the processing time, we have to schedule this job in each of those active slots, including \(C_k\). 
\item Similarly, if \(q>p\), the job corresponding to \(x_{a_q}\) has release time \(C_k\), deadline \(x_{a_q}\), and processing time \(q-p\). Since we assumed that \(x_{a_q}\) is inactive, we have exactly \(q-p-1\) active time slots from variable gadgets and 1 active time slot at \(C_k\) in the interval. As the number of active slots in the job’s interval equals the processing time, we have to schedule this job in each of those active slots, including \(C_k\).
\end{itemize}

Thus, all the \(n_k\) large jobs in the clause gadget have to use the \(C_k\) time slot. In total, this slot is then used by \(n\) jobs from copy gadgets, \(n_k\) jobs from the clause gadget and \(b-n-n_k+1\) unit interval jobs. However this total exceeds \(b\). This is the required contradiction.

We have now proven that all the clauses in the SAT formula are satisfied, as well as that exactly half of the variables were set to true. That completes the proof of Theorem \ref{Theorem:np-complete}.

\section{Conclusion}
In this paper, we showed that the active time scheduling problem is NP-complete when the batch size $b$ is a part of the input. It remains open whether the problem continues to be hard even for constant batch sizes. Another interesting question for future work is whether the known 2-approximations~\cite{chang2017lp,kumar2018brief,calinescu2021new} are best possible, and whether one can show a tight hardness of approximation.

\bibliography{mybibfile}

\end{document}